%% file: StePar2astroph.tex
\documentclass{aa}  

\usepackage{graphicx}
\usepackage{xcolor} 
\usepackage{txfonts}
\bibpunct{(}{)}{;}{a}{}{,}

\begin{document}

\title{{\scshape StePar}: an automatic code to infer stellar atmospheric parameters}
\subtitle{}

\author{H. M. Tabernero\inst{1}
\and
E. Marfil\inst{2}
\and 
D. Montes\inst{2}
\and 
J.~I. Gonz{\'a}lez Hern{\'a}ndez\inst{3,4}
}

   \institute{Centro de Astrobiolog{\'i}a (CSIC-INTA), Carretera de Ajalvir km 4, Torrej{\'o}n de Ardoz, 28850 Madrid, Spain\\
              \email{htabernero@cab.inta-csic.es}
              \and
              Departamento de F{\'i}sica de la Tierra y Astrof{\'i}sica \& IPARCOS-UCM 
              (Instituto de F\'{i}sica de Part\'{i}culas y del Cosmos de la UCM), 
              Facultad de Ciencias F{\'i}sicas, Universidad Complutense de Madrid, 28040 Madrid, Spain
              \and
              Instituto de Astrof\'{i}sica de Canarias (IAC), 38205 La Laguna, Tenerife, Spain
              \and
              Universidad de La Laguna (ULL), Departamento de Astrof\'{i}sica, 38206 La Laguna, Tenerife, Spain
              }

   \date{Received 14 March 2019 / Accepted dd mm 2019}

 
  \abstract
{{\scshape StePar} is an automatic code written in Python 3.X designed to compute the stellar atmospheric parameters $T_{\rm eff}$, $\log{g}$, [Fe/H], and $\xi$ of FGK-type stars by means of the $EW$ method. This code has already been extensively tested in different spectroscopic studies of FGK-type stars with several spectrographs and against myriads of Gaia-ESO Survey UVES U580 spectra of late-type, low-mass stars as one of its thirteen pipelines.}
  {We describe the code and test it against a library of well characterised Gaia benchmark stars. We also release the code to the community and provide the link for download.}
  {We carried out the required $EW$ determination of $\ion{Fe}{i}$ and $\ion{Fe}{ii}$ spectral lines using the automatic tool TAME. {\scshape StePar} implements a grid of MARCS model atmospheres and the MOOG radiative transfer code to compute stellar atmospheric parameters by means of a Downhill Simplex minimisation algorithm.}
  {We show the results of the benchmark star test and also discuss the limitations of the $EW$ method, and hence the code. In addition, we found a small internal scatter for the benchmark stars of 9~$\pm$~32 K in $T_{\rm eff}$, 0.00~$\pm$~0.07 dex in $\log{g}$, and 0.00~$\pm$~0.03 dex in [Fe/H]. Finally, we advise against using {\sc StePar} on double-lined spectroscopic binaries or spectra with R~<~30,000, ${\rm SNR}~<~20$ or $v\sin{i}~>15~{\rm km~s}^{-1}$ as well as stars later than K4 or earlier than F6.}
  {}

   \maketitle

\section{Introduction}

The characterization of stellar spectra is a matter of utmost importance to several fields in modern astrophysics. It provides for the study and better understanding of the different constituents of our galaxy in terms of both individual and large-scale properties of target objects. 

For this reason, stellar spectroscopy stands as a powerful tool that is being widely used in observational surveys, both ongoing and underway, such as the APO Galactic Evolution Experiment \citep[APOGEE,][]{ahn13}, the GALactic Archeology with HERMES \citep[GALAH,][]{sil15}, the LAMOST Experiment for Galactic Understanding and Exploration \citep[LEGUE,][]{den12}, the RAdial Velocity Experiment \citep[RAVE,][]{kun17}, the Sloan Extension for Galactic Understanding and Exploration \citep[SEGUE,][]{lee08}, the Gaia-ESO Survey \citep[GES,][]{gil12}, and WEAVE \citep{dal18}. 

Such large surveys have especially been designed to yield full sets of stellar parameters for as many stars as possible by means of automated methods that ensure the homogeneity of the results. These parameters include the effective temperature, $T_{\rm eff}$; the surface gravity, $\log{g}$; the metallicity, [M/H]; and the micro-turbulent velocity, $\xi$.

In this regard, late-type, low-mass stars of FGK spectral types remain one of the most interesting targets on account of their ubiquity. Furthermore, the optical spectra of these stars have plenty of iron features that are very sensitive to the stellar atmospheric parameters.

The computation of stellar atmospheric parameters of FGK stars under spectroscopic scrutiny is often carried out by means of two different methods: the spectral synthesis method and the $EW$ method. The former uses theoretical synthetic spectra in order to find the best match to a target observed spectrum, whereas the latter uses the strength of several spectral lines to find the set of stellar atmospheric parameters that best reproduce the measured $EW$s. Thorough, recent reviews of these techniques can be found in \citet{all16}, \citet{nis18}, \citet{jof18}, and \citet{bla19}.

There are many implementations of these two methods that are publicly available to the community. Among the spectral synthesis implementations stand the APOGEE pipeline \citep[ASCAP,][]{gar16}, and Spectroscopy Made Easy \citep[SME,][]{pis17, val96}, whereas the $EW$ method is implemented in tools such as FAMA \citep{mag13}, GALA \citep{muc13}, BACCHUS \citep{mas16}, and SPECIES \citep{sot18}. Remarkably, there are also general-purpose toolkits such as the integrated Spectroscopic framework \citep[iSpec, see][]{bla14b} and FASMA \citep{and17, tsa18} that can compute the stellar atmospheric parameters of any given star under both methods.

In general, spectral synthesis methods are based on a $\chi^2$ minimization algorithm based on a precomputed grid of atmospheric models \citep[see][]{val96, gar16, tsa18}. The theoretical spectra, which may sometimes be split up into spectral regions of interest \citep[see e.g.][]{tsa14}, are finally compared with the observations to find the atmospheric model that best fits the data. This approach can also be found in \citet{gon04} and \citet{all06}.

On the other hand, the $EW$ method employs the standard technique based on the iron ionization and excitation balance, taking advantage of the high sensitivity of the strength (i.e. the $EW$) of \ion{Fe}{i} and \ion{Fe}{ii} lines to the variation of the stellar atmospheric parameters. This approach rests on the curves of growth that link, by means of the Saha and Boltzmann equations, the observed $EW$ to the column density of the chemical species that causes the line in the stellar spectrum. Further details on these two equations can be found in e.g. \citet{hub14}. This method has already been applied to several studies found in the literature \citep[see e.g.][]{ghe10, san04, sou08}.

The required $EW$ determination of the Fe lines can be carried out either automatically or manually. There are some automatic tools designed for this task, such as ARES \citep{sou07}, DAOSPEC \citep{ste08}, and TAME \citep{kang12}. All these tools accept some input parameters that can be fine-tuned depending on the quality of the target spectrum under analysis (i.e. the position of the stellar pseudo-continuum, the list of spectral lines to be measured, the parameters that constrain the detection of spectral lines according to the spectral resolution, and so on). In this regard, we highlight the fact that any given linelist is generally assembled from the analysis of a template star \citep[usually the Sun, e.g.][]{san04, sou08}. However, in some cases the template star may be different, e.g. a cool, K-type star \citep{tsa13}, or a giant star \citep{hek07}. The selected lines must be as unblended as possible to avoid the contamination of neighbouring lines that could potentially affect the $EW$ measurements.

In this work we present a full description of the automatic code {\scshape StePar}, written in Python 3.X, which is based on the $EW$ method. This code has already been applied to the careful study of FGK-type stars \citep{gon12, tab12, tab14, tab17, jof17, mon18} and has also been extensively used to automatically analyse hundreds of Gaia-ESO UVES spectra since it stands as one of the thirteen pipelines that characterises the UVES U580 spectra of late-type, low-mass stars \citep[see][]{smi14,lan15}.  

A concise description of the complete method is found in Sect.~\ref{sect_ste}. Detailed explanations of the {\scshape StePar} internal workflow are given in Sect.~\ref{sect_test}. In Sect.~\ref{sect_dis} we discuss our results and compare them to those obtained in previous works, as well as to evolutionary tracks. Finally, in Sect.~\ref{sect_con} we warn of the limitations of the $EW$ method, and hence {\scshape StePar}.

\section{The {\scshape StePar} code}
\label{sect_ste}

The basic ingredients that {\scshape StePar} needs to derive the stellar atmospheric parameters $T_{\rm eff}$, $\log{g}$, [Fe/H], and $\xi$, fully explained in Sect.~\ref{steparworkflow}, are the following:
\begin{itemize}
	\item[(i)] A grid of stellar atmospheric models: MARCS model atmospheres \citep{gus08}.
	\item[(ii)] A code to solve the radiative transfer problem under the assumption of local thermodynamic equilibrium (LTE): the MOOG code \citep{sne73}.
	\item[(iii)] A list of $\ion{Fe}{i}$ and $\ion{Fe}{ii}$ spectral lines along with their atomic parameters.
	\item[(iv)] A programme to measure the required $EW$s for later use: TAME \citep[][]{kang12}. 
    \item[(v)] An optimization algorithm: the Downhill Simplex method \citep{pre02}.
\end{itemize}

\subsection{{\scshape StePar} workflow}
\label{steparworkflow}

\begin{figure}
    \centering
    \includegraphics[scale = 0.9]{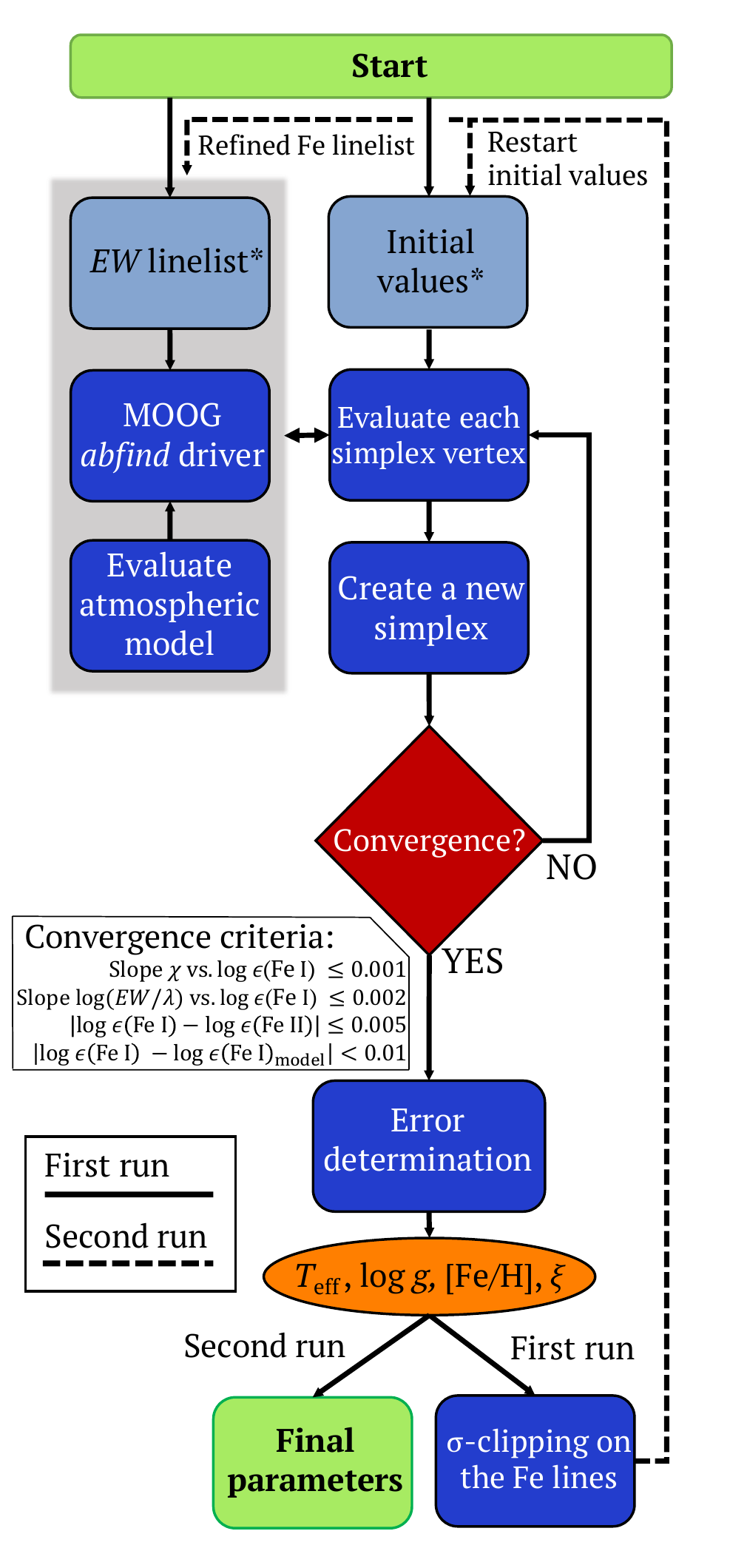}
    \caption{{\scshape StePar} workflow diagram. As explained in Sect.~\ref{steparworkflow}, for any given star under analysis, the code performs two simplex runs. During the second simplex run, the initial values, which are initially set to the solar canonical values, are reset to the values obtained during the first run. Likewise, the $EW$ linelist is refined according to the $\sigma$-clipping procedure on the $\ion{Fe}{i}$ lines. After the second run, the code halts execution and yields the final solution for the star.}
    \label{fig:workflow}
\end{figure}

The stellar atmospheric parameters of FGK-type stars, namely $T_{\rm eff}$, $\log{g}$, $\xi$, and [Fe/H], can be derived in an automated fashion with {\scshape StePar}\footnote{{\scshape StePar} is available at \url{https://github.com/hmtabernero/StePar} under the two-clause BSD license.}. Its workflow is shown in Fig.~\ref{fig:workflow}. In the standard {\scshape StePar} version presented here, we employed the 2017 version of the MOOG code \citep[via the {\tt abfind} driver, see][]{sne73} and a grid of plane-parallel and spherical MARCS model atmospheres \citep{gus08}, although other model grids can be used alongside {\scshape StePar} \citep[see][]{tab12,tab17,mon18}. For lower gravities ($\log{g} <$ 3.5) we used the spherical grid, whereas we employed the non-spherical grid  for greater gravities ($\log{g} \geq$ 3.5). Although MOOG treats the MARCS spherical atmospheric models as if they were plane-parallel, \citet{hei06} proved that this potential inconsistency is negligible.  However, since the MARCS grid is finite, {\scshape StePar} includes an interpolation subroutine, based on the Python Scipy library, which draws on prior knowledge of the desired model and its neighbouring grid models to interpolate between them \citep{bar96}.

{\scshape StePar} needs a MOOG-compliant $EW$ file as input, which can be provided by the user in the proper format using an automatic measurement tool. This MOOG-input file must contain the following atomic data for each line considered in the analysis: 

\begin{itemize}
\item[(i)] Central wavelength of the line, in \AA;
\item[(ii)] A number that indicates the atomic number and ionisation stage of the chemical species that causes the line (26.0 and 26.1 in the case of \ion{Fe}{i} and \ion{Fe}{ii} lines, respectively);
\item[(iii)] the excitation potential, $\chi$, in eV;
\item[(iv)] the oscillator strength, $\log{gf}$;
\item[(v)] the $EW$ of the line in m\AA.
\end{itemize}

To perform our analysis we opted for the automatic code TAME\footnote{TAME can be downloaded from \url{http://astro.snu.ac.kr/~wskang/tame/}} \citep{kang12}, which can be run in either an automated or manual mode. Its manual mode has an interface that allows some user control over the $EW$ measurements to check problematic spectra when needed. We followed the approach of \citet{kang12} to adjust the {\tt rejt} parameter of TAME according to the $S/N$ of each spectrum. The other TAME parameters we employed were:

\begin{itemize}
\item[(i)]{\tt smoother} = 4, the recommended parameter for smoothing the derivatives used for line identification;
\item[(ii)]{\tt space} = 3, the wavelength interval, in \AA, from each side of the central line to perform the $EW$ computation;
\item[(iii)]{\tt lineresol} = 0.1, the minimum distance, in \AA, between two lines for TAME to resolve them;
\item[(iv)]{\tt miniline} = 2, the minimum $EW$ that will be printed in the output. 
\end{itemize}

Further details on TAME parameters can be found in \citet{kang12}. In addition, we only considered measured lines with 10 m\AA~$<$~$EW$~$<$~120~m\AA~to avoid problems with line profiles of very intense lines and tentatively bad $EW$ measurements of extremely weak lines. For benchmark stars that have two high signal-to-noise spectra available, we applied an additional filter on the $\ion{Fe}{i, ii}$ lines and rejected the ones that might have a differential equivalent width beyond three times the standard deviation of the $EW$ differences between the corresponding $\ion{Fe}{i, ii}$ lines measured on each of those two spectra.

As damping prescription, we used the Anstee-Barcklem-O'Mara \citep[ABO, see][]{abo98} data (if available), through option 1 of MOOG. The atmospheric parameters can then be inferred from previously assembled $\ion{Fe}{i}$-$\ion{Fe}{ii}$ linelists.

The minimisation procedure of {\scshape StePar} is the Downhill Simplex algorithm \citep{pre02}, which tries to minimise a quadratic form composed of the excitation and ionization equilibrium conditions to find the best parameters of the target star. This minimisation algorithm can reach convergence in very few iterations, and it is so fast that it is, for instance, the optimization method of choice for the ASCAP pipeline in APOGEE \citep{ahn13}. Since it does not use derivatives, they have to be estimated numerically, such as in the Levenberg-Marquardt method.  If we let $\log{\epsilon}$(\ion{Fe}{i}), and $\log{\epsilon}$(\ion{Fe}{ii}) stand for the Fe abundance returned by the \ion{Fe}{i} and \ion{Fe}{ii} lines, respectively, and $\log{(EW/\lambda)}$ be their reduced equivalent width, {\scshape StePar} iterates until the slopes of $\chi$ versus $\log{\epsilon(\textrm{\ion{Fe}{i}})}$ and $\log{(EW / \lambda)}$ versus $\log{\epsilon(\textrm{\ion{Fe}{i}})}$ are virtually zero, i.e. excitation equilibrium, and imposing ionization equilibrium, so that $\log{\epsilon(\textrm{\ion{Fe}{i}})} = \log{\epsilon(\textrm{\ion{Fe}{ii}})}$. Throughout this procedure, we checked that the [Fe/H] obtained from the iron lines is always compatible with the metallicity of the input atmospheric model. The actual convergence criteria of {\sc StePar}, as shown in Fig.~\ref{fig:workflow}, are the following:
\begin{itemize}
\item[(i)]{Slope $\xi$ vs. $\log{\epsilon(\textrm{\ion{Fe}{i}})} \le 0.001$;}
\item[(ii)]{Slope $\log{EW/\lambda}$ vs. $\log{\epsilon(\textrm{\ion{Fe}{i}})} \le 0.002$;}
\item[(iii)]{$|\log{\epsilon(\textrm{\ion{Fe}{i}})} - \log{\epsilon(\textrm{\ion{Fe}{ii}})}|$ $\le 0.005$;}
\item[(iv)]{$|\log{\epsilon(\textrm{\ion{Fe}{i}})} - \log{\epsilon(\textrm{\ion{Fe}{i}})_{\rm model}}| < 0.01$.}
\end{itemize}

\begin{figure}
    \includegraphics[width=0.48\textwidth]{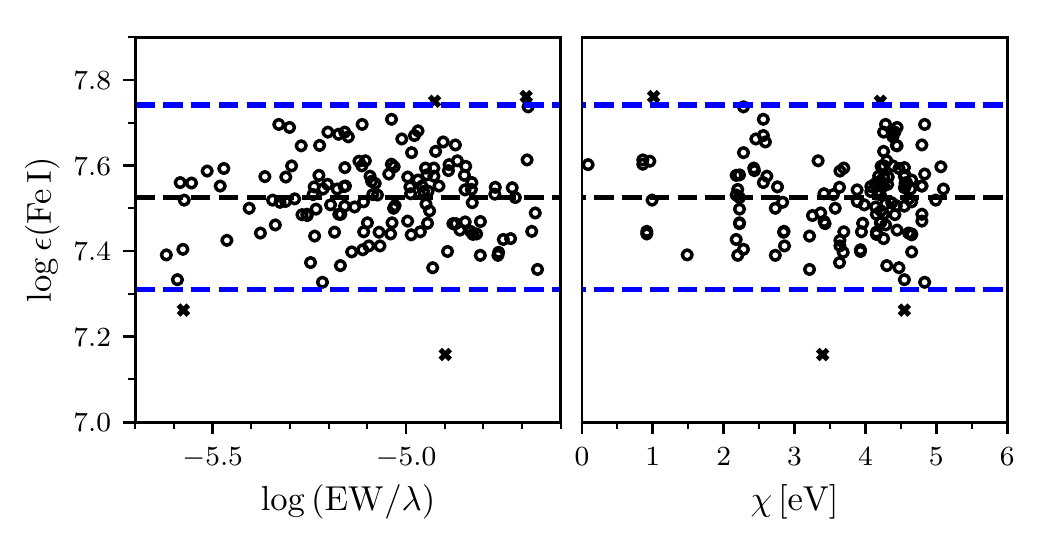}
    \caption{{\scshape StePar} inner $3\sigma$-clipping of the $\ion{Fe}
             {i}$ lines on
             the NARVAL spectrum of the Sun. $\log{\epsilon}$(\ion{Fe}{i}) stands for the Fe abundance returned by the Fe lines while $\log{({\rm EW}/\lambda)}$ represents their reduced $EW$s. Black crosses depict the rejected \ion{Fe}{i} lines. The dashed black lines represent a linear fit to the points, whereas the dashed blue lines are located at the $3\sigma$-level.}
    \label{fig:sigmaclipping}
\end{figure}

For each target spectrum, {\scshape StePar} performs two simplex runs, which in turn individually entail a full parameter determination using the Downhill Simplex optimization method. The first run deals with the $EW$s file as initially measured by TAME. Next, the best model that is found in this first run by the optimization routine is evaluated. {\scshape StePar} then performs a $3\sigma$-clipping procedure on the $\ion{Fe}{i}$ abundance values obtained from the $\ion{Fe}{i}$ lines, so that we can remove the outliers, if any, due to "wrong" $EW$s measurements that could potentially invalidate the analysis. The second run is finally launched on a new input $EW$ file that does not contain the rejected lines. An example of this $3\sigma$-clipping procedure is shown in Fig.~\ref{fig:sigmaclipping}. As to execution time, {\scshape StePar} takes between 2 and 5 minutes to perform this whole procedure per star, depending on its actual position in the FGK parameter space.

To feed this minimisation process, the canonical solar values are used as initial input values ($T_{\rm eff}=5777$ K, $\log{g}=4.44$ dex, $\xi$~=~1~$~{\rm km~s}^{-1}$). However, {\scshape StePar} is able to reach a solution even if the problem star is different from the Sun (e.g. a metal-poor giant). In other words, the final solution for any given star is independent of the initial set of parameters employed.

Finally, the uncertainties in the stellar parameters are determined as follows: 
\begin{itemize}
	\item[(i)] For the micro-turbulence, we slightly change the value of $\xi$ until the slope of $\log{\epsilon(\textrm{\ion{Fe}{i}})}$ vs. $\log{(EW / \lambda)}$ 
	           varies within its own error, divided by the 
	           square root of the number of $\ion{Fe}{i}$ lines. 
	\item[(ii)] The effective temperature is varied until the slope of $
	            \log{\epsilon(\textrm{\ion{Fe}{i}})}$ vs. $\chi$
	            increases up to the error on the slope, divided by the square root of the 
	            number of $\ion{Fe}{i}$ lines. 
	            By increasing $\xi$ on its error, we recompute the effective 
	            temperature. These two sources of error are added in quadrature.
	\item[(iii)] The surface gravity is then varied until the $\ion{Fe}{ii}$ 
	             abundance increases by a quantity 
	             equal to the standard deviation divided by the square root of the 
	             number of $\ion{Fe}{ii}$ lines. 
	             All the previous errors in $\xi$ and $T_{\rm eff}$ are taken 
	             into account by varying these
	             quantities separately, thus recomputing the gravity. These 
	             differences are later added in quadrature.
	\item[(iv)] Finally, to determine the error in the Fe abundance, the
	            stellar atmospheric parameters are varied in their respective 
	            uncertainties, 
	            which enables the combination of all the $\ion{Fe}{i, ii}$ variations due 
	            to the stellar parameters uncertainties and the standard deviation 
	            of the $\ion{Fe}{i, ii}$ abundances in quadrature.
\end{itemize}

\section{Testing the code}

\label{sect_test}
\subsection{Selection of the $\ion{Fe}{i, ii}$ linelists}

The $EW$ method requires a significantly large selection of reliable $\ion{Fe}{i}$ and $\ion{Fe}{ii}$ lines. In principle, reliable lines are meant to be clean spectral lines that are not strongly affected by line blending of neighbouring lines or conspicuous spectral features. All the same, the available atomic data of these clean lines may not be precise enough for a trustworthy analysis. The main problem stems from the tabulated values of the transition probability per unit time of the spectral lines, $\log{gf}$. Some authors avoid this problem by calibrating this value for each line \citep[see e.g.][]{san04,sou08,nev09}, normally by means of an inverse solar analysis, in which they vary the $\log{gf}$ value for a given line until they recover the corresponding solar abundance value. These are called the astrophysical $\log{gf}$s. 

The Gaia-ESO linelist \citep{hei15b} was originally extracted from a variety of sources with the aim of finding the best atomic parameters available. It contains around 560 $\ion{Fe}{i, ii}$ features whose parameters were mostly taken from the Vienna Atomic Line Database \citep[VALD\footnote{\url{http://www.astro.uu.se/~vald/}},][]{rad15}. The prime goal was to compile a trustworthy selection of lines to compute high-precision stellar parameters.

However, given the diversity of stars across the Milky Way (metal-poor dwarfs and giants, solar-type stars, metal rich giants, etc.) in the Gaia-ESO Survey, it soon became apparent that one linelist may fall short for the analysis of any given star. Hence, the analysis of any stellar sample under {\scshape StePar} is set to rely on four template stars from which four different lists of $\ion{Fe}{i, ii}$ lines are assembled: the Sun, HD~22879, $\xi$~Hya, and Arcturus. The corresponding linelists of \ion{Fe}{i} and \ion{Fe}{ii} can be found in Table~\ref{tab:line_table_all_fe_i} and Table~\ref{tab:line_table_all_fe_ii}, respectively. These four template stars, which fully cover the FGK parameter space, as explained below, help us classify any star prior to the analysis with {\scshape StePar}.

\begin{table}
\caption{Linelist template stars and their reference stellar atmospheric parameters from \citet{hei15a}, with updated values from \citet{jof18}.}
\label{tablasol2}
\centering
\begin{tabular}{llcccc}
\hline\hline\noalign{\smallskip}
Star      & List & $T_{\rm eff}$ [K] & $\log{g}$ [dex] & [Fe/H] [dex]       \\
\hline\noalign{\smallskip}
Sun       & MRD  & 5777  $\pm$ 1        & 4.44 $\pm$ 0.01 &       0.03 $\pm$ 0.01 \\
HD~22879  & MPD  & 5868  $\pm$ 89       &  4.27 $\pm$ 0.03 &    $-$0.86 $\pm$ 0.05 \\
$\xi$~Hya & MRG  & 5044  $\pm$ 40 & 2.87 $\pm$ 0.02 & 0.16 $\pm$ 0.20 \\
Arcturus  & MPG  & 4286  $\pm$ 35       &  1.60 $\pm$ 0.20 &    $-$0.52 $\pm$ 0.08 \\
\hline
\end{tabular}
\end{table}

Such division of the parameter space meets the following criteria. In terms of metallicity, we distinguish between metal-rich stars, i.~e. [Fe/H] $\geq$ $-$0.30, and metal-poor stars, i.~e. $-$0.30 $>$ [Fe/H] $\geq$ $-$1.50. In terms of surface gravity, we make a distinction between the giant regime, i.~e. $\log{g}$ $<$ 4.00, and the dwarf regime, i.~e. $\log{g} \geq$ 4.00. These partitions mean that the global parameter space is divided into four different regions: metal-rich dwarfs (MRD), metal-poor dwarfs (MPD), metal-rich giants (MRG), and metal-poor giants (MPG). Because of their scarcity, we decided to set aside the extremely metal-poor stars with [Fe/H] $<$ $-$1.50. The general scheme of this division (MRD, MPD, MRG, MPG) is shown in Table~\ref{tablasol2}.

As already mentioned, the analysis of any given star is done blindly, that is, its stellar parameters are not known beforehand. Hence, it is not known a priori which linelist corresponds to any given star. In order to overcome this issue, we measured the lines from all four linelists and did a first-pass with {\scshape StePar} so we could finally assign a linelist to each star depending on the parameters obtained. This preliminary step allowed us to run {\scshape StePar} with the corresponding linelist to get the final solution for the star.

\subsection{Gaia benchmark test}

Every spectroscopic study requires a reference point to assess the validity of the obtained results. Despite the Sun being widely used in the literature as a common reference, the use of one single star as a central point of reference provides no evidence on how a given method works in different regions of the parameter space.

In this regard, the Gaia benchmark stars were originally meant as calibrators to test the different approaches to the analysis. The availability of specific information for these stars was the key to determining their stellar atmospheric parameters independently from spectroscopy \citep{hei15a}. In this sense, the Gaia benchmark stars represent a cornerstone when it comes to weighing the impact of the general limitations (e.g. wavelength coverage, linelists employed, resolution, etc.) inherent to any spectroscopic method that aims at the computation of such parameters.

Ideally, with a common background, any set of tools using the same data should converge to the same atmospheric parameters. However, the fact that every method takes into account a different set of spectral lines inevitably leads to slightly different stellar parameters. Although, in general, these differences are mostly dependent on the radiative transfer code, the stellar atmospheric models, the specific method, and the input data, among the GES nodes this dependence mostly comes down to the input data (i.e. measurement of $EW$s, signal-to-noise ratio of the spectra, the local continuum normalisation, etc.) and the method that each node takes into consideration.

These stars, 23 in total, were taken from the stellar library\footnote{The spectra of these stars can be downloaded from \url{https://www.blancocuaresma.com/s/benchmarkstars}} described in \citet{bla14a}, which covers the optical region 4800--6800~\AA. The stars span a wide range in the parameter space \citep{hei15a, jof15} that allowed us to test the reliability of the results given by {\scshape StePar} (see Table~\ref{tab:reference_parameters}). The internal consistency of the code was tested by deriving stellar atmospheric parameters from one or two high signal-to-noise spectra of the same star (see Table~\ref{tab:stepar_results}).

\begin{table}
\centering
\caption{\label{tab:nlines} Number of \ion{Fe}{i} and \ion{Fe}{ii} in each of the four linelists used in this work. The wavelength coverage of all four lists is 4800--6800~\AA, in line with that of the spectra under analysis. We also display the number of stars in each category.}
\begin{tabular}{lcccc}
\hline\hline\noalign{\smallskip}
Element      & MRD & MPD & MRG & MPG \\
\hline\noalign{\smallskip}
\ion{Fe}{i}  & 146 & 127 & 113 & 115 \\
\ion{Fe}{ii} & 12  & 13  & 11  & 6   \\
\hline
\#stars &  8 & 4  & 7 & 4 \\
\hline
\end{tabular}
\end{table}

\begin{figure*}
    \includegraphics[width=\textwidth]{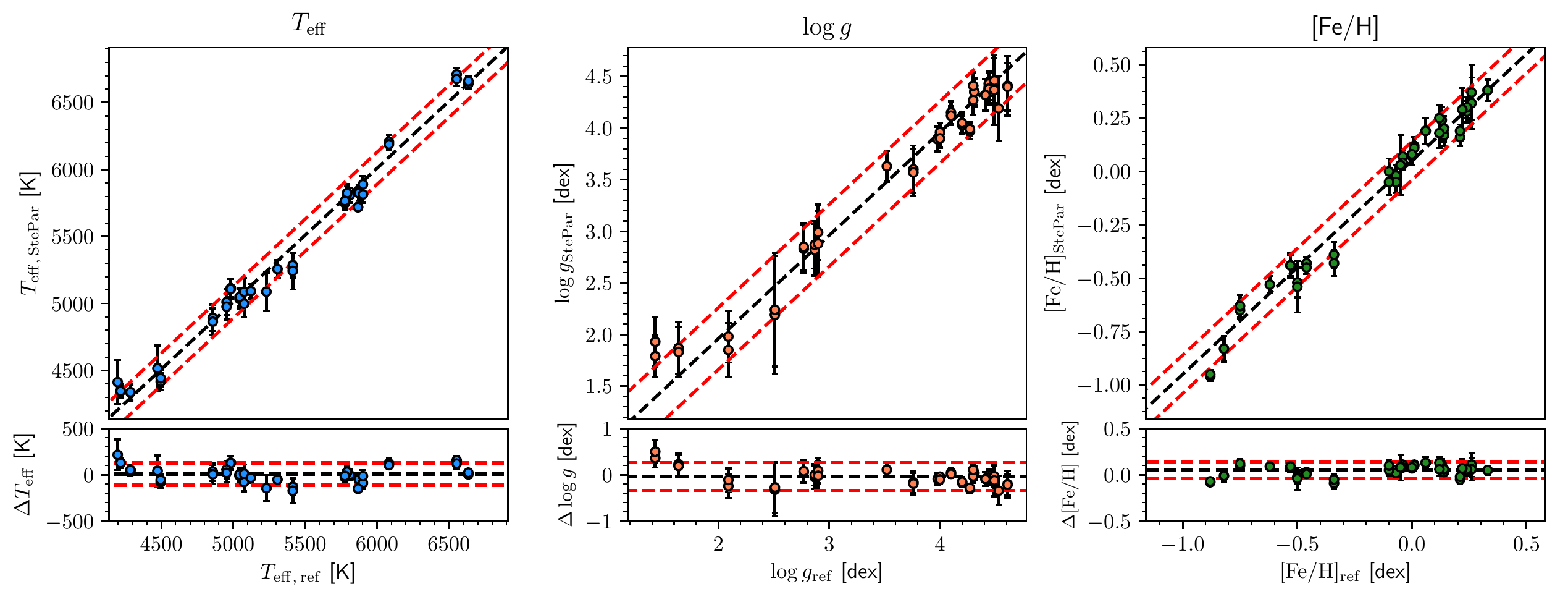}
    \caption{{\scshape StePar} results for the Gaia benchmark stars plotted against the literature values taken from \citet{hei15a},  with updated values from \citet{jof18}. The upper panels show a one-to-one correspondence whereas the bottom panels depict the absolute differences.  According to the values shown in Tab.~\ref{table_mon}, dashed black lines in the upper panels correspond to a one-to-one relationship, shifted following the average differences in each parameter, whereas in the bottom panels, they are centered at the average differences. The dashed red lines in all panels correspond to a margin of 120 K, 0.30 dex and 0.09 dex in $T_{\rm eff}$, $\log{g}$ and [Fe/H], respectively, according to the $\sigma$ values found in the differences in each parameter.}
    \label{figpar}
\end{figure*}

\begin{figure}
\includegraphics[width=0.48\textwidth]{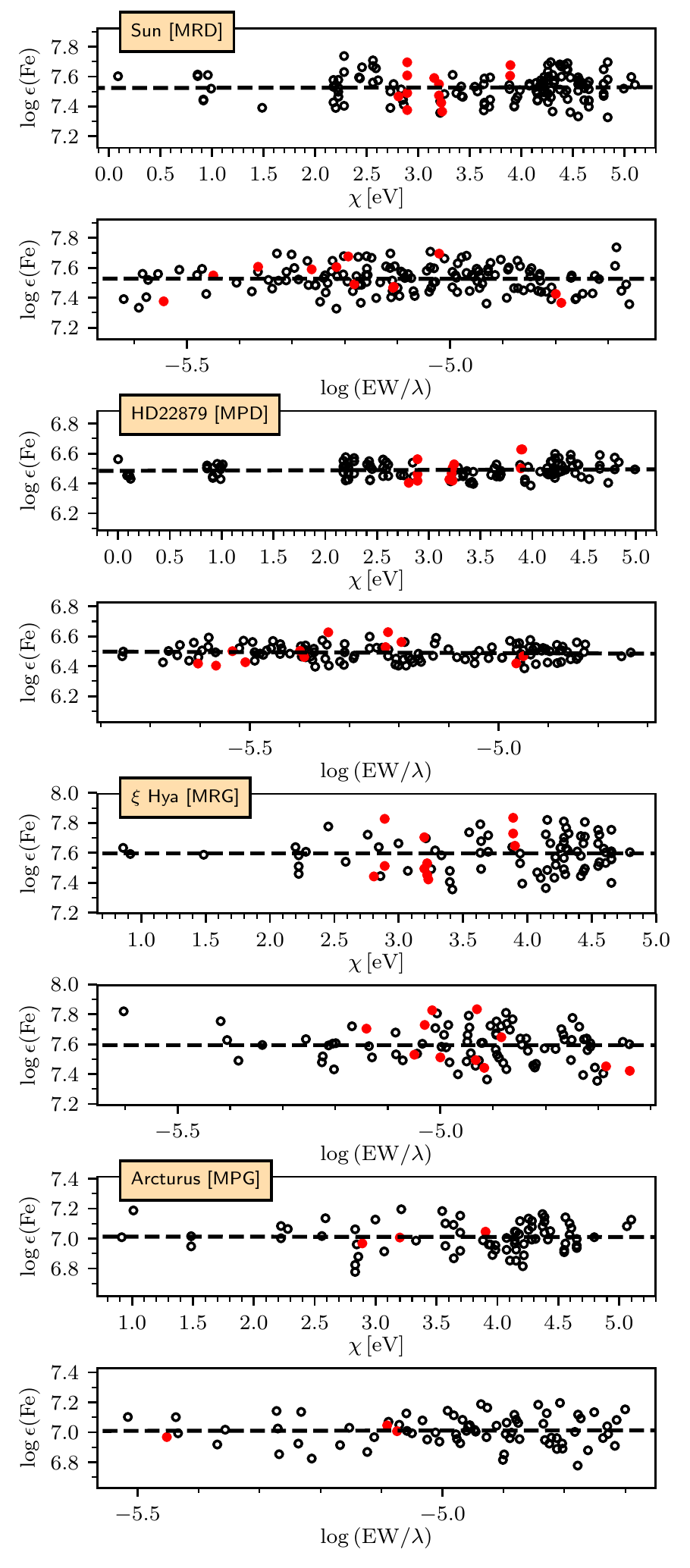}
 \caption{From top to bottom: line iron abundance retrieved by {\scshape StePar} for the final solution of the four reference stars: the Sun (NARVAL), HD~22879 (NARVAL), $\xi$~Hya (ESPaDOns) and Arcturus (UVES).  $\log{\epsilon}$(\ion{Fe}{i}) stands for the Fe abundance returned by the Fe lines while $\log{({\rm EW}/\lambda)}$ represents their reduced EWs. Unfilled black dots represent \ion{Fe}{i} lines, whereas red dots depict \ion{Fe}{ii} lines. The dashed black lines represent the least-squares fit to the data points.}
\label{fig:StePar4ref}
\end{figure}

\section{Discussion}
\label{sect_dis}

\begin{table}
\caption{Summary of the Montecarlo simulations performed using the stellar atmospheric parameters calculated in Sect.~\ref{sect_test}. Hereby we present the average difference on each parameter, along with the values of the Pearson ($r_{\rm p}$) and the Spearman ($r_{\rm S}$) correlation coefficients.}
 \label{table_mon}
 \centering
 \begin{tabular}{lccc}
 \hline\hline\noalign{\smallskip}
 Parameter            & Difference                & $r_{\rm p}$        & $r_{\rm S}$       \\
 \hline\noalign{\smallskip}
 $T_{\rm eff}$ [K]    &    9~$\pm$~120      & $-$0.12~$\pm$~0.10 & $-$0.15~$\pm$~0.10\\
 $\log{g}$ [dex]      &    $-$0.04~$\pm$~0.30 & $-$0.33~$\pm$~0.16 & $-$0.26~$\pm$~0.14\\
 {[Fe/H]} {[dex]}     &    0.05~$\pm$~0.09  &    0.25~$\pm$~0.10 &    0.16~$\pm$~0.11\\
 \hline
\end{tabular}
\end{table}

\begin{figure}
  \includegraphics[width=0.5\textwidth]{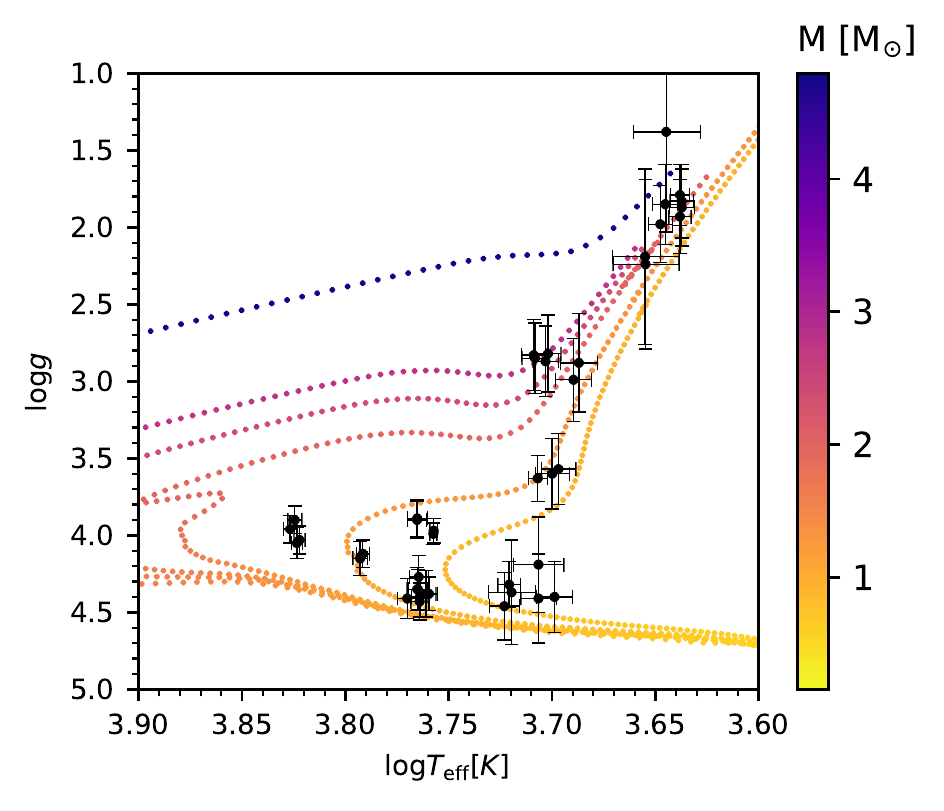}
  \caption{Kiel diagram ($\log{g}$ vs. $\log{T_{\rm eff}}$) for all the spectra alongside the YaPSI isochrones for 0.1, 0.4, 0.6, 1, 4, and 13 Ga \cite[for Z=0.016, see][]
  {spa17}.}
  \label{fig:kieldiag}
  \end{figure}

Although different spectra of the same object should theoretically result in exactly the same stellar atmospheric parameters, we noticed slight deviations in our analysis. The mean differences for the benchmark stars are 9~$\pm$~32 K in $T_{\rm eff}$, 0.00~$\pm$~0.07 dex in $\log{g}$, and 0.00~$\pm$~0.03 dex in [Fe/H].

These internal differences are mostly due to the quality of the individual spectra. The average uncertainties for the benchmark stars are 75 K in $T_{\rm eff}$, 0.21 dex in $\log{g}$, and 0.06 dex in [Fe/H]. Furthermore, the average uncertainties are greater than the scatter that arises from the analysis of different spectra of the same object, as expected. Other sources of uncertainty might arise from systematic effects inherent to any methodology.

We compared the values of the stellar parameters for each spectrum to the reference values, as shown in Fig.~\ref{figpar}. We found the following differences, all of which might be systematic: $\Delta T_{\rm eff} = 9\pm89$ K, $\Delta\log{g}=-0.04\pm0.18$ dex, and $\Delta$[Fe/H] = 0.05~$\pm$~0.06 dex. At first glance, there are no notable differences at the 1-2$\sigma$~level. Additionally, in Fig.~\ref{fig:StePar4ref} we plot the line iron abundance retrieved by {\scshape StePar} for the final solution of the four reference stars.

However, since systematic trends may still remain hidden, we performed 10,000 Monte Carlo simulations on our data \citep[as in][]{tab18} in the hope of assessing possible sources of tentatively systematic offsets. We took each atmospheric parameter ($T_{\rm eff}$, $\log{g}$, and [Fe/H]) to check any possible correlations by means of the Pearson and Spearman correlation coefficients, which are a measure of the correlation between any two given variables. Specifically, the Pearson coefficient is a classical correlation indicator, whereas the Spearman coefficient is a more robust non-parametric estimator of the statistical dependence of any two given variables \citep[for more details, see e.g.][]{pre02}. In Table~\ref{table_mon} we clearly show that no systematic trends are present above the $2\sigma$ level (i.e. within 95\% confidence interval). These results are shown in Fig.~\ref{figpar} and Fig.~\ref{fig:maspar}. Interestingly enough, even if the offsets in effective temperature and surface gravity are noticeable, the offset in the iron abundance is negligible (see Fig.~\ref{figpar}). This is probably due to the fact that {\scshape StePar} produces self-consistent stellar parameters. For example, a deviation on the effective temperature can be compensated by the other parameters.

\begin{table}
\caption{Same differences as shown in Table~\ref{table_mon} organised on a per-linelist basis, i.e. MRD, MPD, MRG, and MPG lists.}
\label{table_type}
\centering
\begin{tabular}{lccc}
\hline\hline\noalign{\smallskip}   
\multicolumn{4}{c}{Difference}\\
\hline\noalign{\smallskip}

List & $T_{\rm eff}$ [K] & $\log{g}$ [dex] & [Fe/H] [dex]\\
\hline\noalign{\smallskip}
MRD  &    14~$\pm$~105   & 0.05~$\pm$~0.19 & 0.06~$\pm$~0.07 \\
MPD  & $-$88~$\pm$~96    & 0.16~$\pm$~0.18 & 0.03~$\pm$~0.06 \\
MRG  &    44~$\pm$~122   & 0.05~$\pm$~0.36 & 0.08~$\pm$~0.08 \\
MPG  &    28~$\pm$~92    & 0.15~$\pm$~0.31 & 0.05~$\pm$~0.09 \\
\hline\noalign{\smallskip}
\end{tabular}
\end{table}

Finally, we assessed the performance of {\scshape StePar} in different regions of the parameter space. In Fig.~\ref{fig:kieldiag} we plot a Kiel diagram, i.e. $\log{g}$ vs. $\log{T_{\rm eff}}$, including the latest state-of-the-art YaPSI isochrones \citep{spa17}. In light of this figure, we do not find any major inconsistencies with the parameter space encompassed by the isochrones in general terms. However, we must notice that our method does not reproduce the gravity of K-stars, as they should have higher values. The former result is not entirely unexpected \citep{tab12}, as this result might be due to an ionization imbalance problem \citep{tsa19}. We still find higher effective temperatures for the F-type dwarfs, although they deviate less than in previous works \citep[i.e.][]{tab17,mon18}. Finally, in Table~\ref{table_type} we show how {\scshape StePar} performs equally well in the four regions of the FGK parameter space defined in this work, i.e., MRD, MPD, MRG, and MPG, although the MRD and MPD linelists produce slightly larger errors than the rest. Interestingly enough, the errors in surface gravity are larger in the giant regime compared with the dwarf regime. As to the metallicity, we find a similar scatter in the four regions. Despite these differences, we find a good agreement with the reference values.

\section{Conclusions}
\label{sect_con}
In this work, we have presented a robust code, {\scshape StePar}, which will hopefully be useful to the community when deriving the stellar atmospheric parameters of FGK-type stars under the $EW$ method. This code has already been tested during the last few years against myriads of Gaia-ESO high-resolution spectra. We have also tested {\scshape StePar} against a library of Gaia Benchmark stars. Although we find some differences with the reference parameters, they are not significant.

Finally, we want to address some general limitations of the $EW$ method that should be taken into consideration when using {\scshape StePar}. First, data must have enough quality to be analysed. This means that, since low signal-to-noise ratios (SNR) may translate into poor $EW$ measurements, we highly recommend placing a cut at SNR~$<$~20. In addition, we underline the importance of placing a lower limit in spectral resolution at R\,=\,30,000 to prevent undesired line blending and suboptimal placement of the continuum level \citep{fre02}. Second, {\scshape StePar} cannot derive stellar atmospheric parameters of fast-rotating stars. Although the $EW$s are not, in fact, altered by rotation, the line profiles are indeed affected by rotational broadening and may no longer fit a Gaussian profile properly. In this sense, blending of neighbouring lines to the one of interest can also make it nearly impossible to get a reliable $EW$ estimate for a given line. In these cases, we advise against using {\scshape StePar} on any star with a rotational velocity higher than 15 km~s$^{-1}$. In addition,  double-lined spectroscopic binaries should also be removed from any sample to avoid obtaining unreliable parameters. Finally, we do not recommend to derive stellar parameters with {\scshape StePar} for stars earlier than F6 and later than K4.

\begin{acknowledgements}
We would like to thank the anonymous referee for their insightful comments and suggestions to improve the manuscript of the paper. The authors acknowledge financial support from the Spanish Ministerio de Ciencia, Innovaci\'{o}n y Universidades through projects AYA2016-79425-C3-1 (UCM), AYA2016-79425-C3-2 (CAB), and AYA2017-86389-P (IAC). E.~M. acknowledges financial support from the Spanish Ministerio de Educaci\'{o}n y Formaci\'{o}n Profesional through fellowship FPU15/01476. J.~I.~G.~H. also acknowledges financial support from the Spanish MINECO (Ministry of Economy of Spain) under the 2013 Ram{\'o}n y Cajal programme MINECO RyC-2013-14875.
\end{acknowledgements}

\bibliographystyle{aa} 
\bibliography{StePar2astroph}

\begin{appendix}
\section{Appendix}
\include{Tables/benchmarks}

\include{Tables/stepar_results_last}
\include{Tables/lines_fe_i}

\include{Tables/lines_fe_ii}

 \begin{figure*}
     \includegraphics{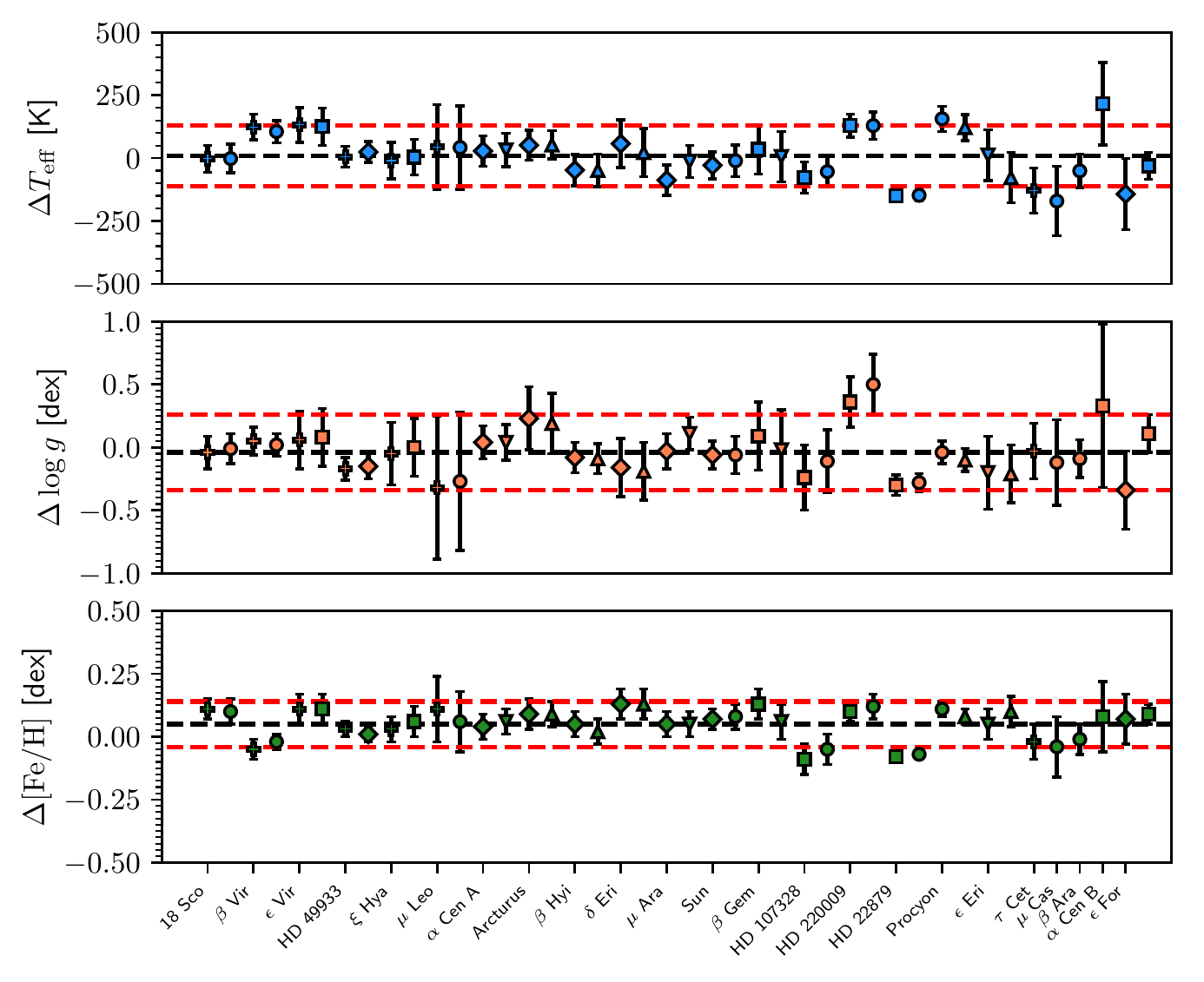}
     \caption{{\scshape StePar} differences with respect to the reference values \citep{hei15a} where each symbol denotes a different spectrograph:  NARVAL (circles). HARPS.GBOG (squares), HARPS.Archive (diamonds), UVES.POP (upward triangles), UVES (downward triangles), ESPaDOnS (crosses).}
     \label{fig:maspar}
 \end{figure*}
 \end{appendix}
\end{document}

%% file: Tables/benchmarks.tex
\longtab[1]{
\begin{longtable}{ccccc}

\caption{\label{tab:reference_parameters} Reference stellar atmospheric parameters of the Gaia benchmark stars taken from \citet{hei15a}, with updated values from \citet{jof18}. }\\
\hline\hline\noalign{\smallskip}
Star & Spectral type & $T_{\rm eff}$ & $\log{g}$ & [Fe/H] \\
     &               & [K]           & [dex]     & [dex]  \\
\noalign{\smallskip}\hline\noalign{\smallskip}
\endfirsthead
\multicolumn{5}{c}{Metal-rich dwarfs (MRD)}\\
\hline\noalign{\smallskip}
Procyon        & F5IV-V   & 6554 $\pm$  84 & 4.00 $\pm$ 0.02 &    0.01 $\pm$ 0.08 \\
$\beta$~Vir    & F9V      & 6083 $\pm$  41 & 4.10 $\pm$ 0.02 &    0.24 $\pm$ 0.07 \\
$\mu$~Ara      & G3IV-V   & 5902 $\pm$  66 & 4.30 $\pm$ 0.03 &    0.35 $\pm$ 0.13 \\
18~Sco         & G2Va     & 5810 $\pm$  80 & 4.44 $\pm$ 0.03 &    0.03 $\pm$ 0.03 \\
$\alpha$~Cen~A & G2V      & 5792 $\pm$  16 & 4.31 $\pm$ 0.01 &    0.26 $\pm$ 0.08 \\
Sun            & G2V      & 5771 $\pm$   1 & 4.44 $\pm$ 0.00 &    0.03 $\pm$ 0.05 \\
$\alpha$~Cen~B & K1V      & 5231 $\pm$  20 & 4.53 $\pm$ 0.03 &    0.22 $\pm$ 0.10 \\
$\epsilon$~Eri & K2Vk:    & 5076 $\pm$  30 & 4.61 $\pm$ 0.03 & $-$0.09 $\pm$ 0.06 \\
\hline\noalign{\smallskip}
\multicolumn{5}{c}{Metal-poor dwarfs (MPD)}\\
\hline\noalign{\smallskip}
HD~49933       & F2V      & 6635 $\pm$  91 & 4.20 $\pm$ 0.03 & $-$0.41 $\pm$ 0.08 \\
HD~22879       & F9V      & 5868 $\pm$  89 & 4.27 $\pm$ 0.04 & $-$0.86 $\pm$ 0.05 \\
$\tau$~Cet     & G8.5V    & 5414 $\pm$  21 & 4.49 $\pm$ 0.02 & $-$0.49 $\pm$ 0.03 \\
$\mu$~Cas      & G5Vb     & 5308 $\pm$  29 & 4.41 $\pm$ 0.06 & $-$0.81 $\pm$ 0.03 \\
\hline\noalign{\smallskip}
\multicolumn{5}{c}{Metal-rich giants (MRG)}\\
\hline\noalign{\smallskip}
$\beta$~Hyi    & G0V      & 5873 $\pm$  45 & 3.98 $\pm$ 0.02 & $-$0.04 $\pm$ 0.06 \\
$\xi$~Hya      & G7III    & 5044 $\pm$  40 & 2.87 $\pm$ 0.02 &    0.16 $\pm$ 0.20 \\
$\epsilon$~Vir & G8III    & 4983 $\pm$  61 & 2.77 $\pm$ 0.02 &    0.15 $\pm$ 0.16 \\
$\delta$~Eri   & K1III-IV & 4954 $\pm$  30 & 3.76 $\pm$ 0.02 &    0.06 $\pm$ 0.05 \\
$\beta$~Gem    & K0IIIb   & 4858 $\pm$  60 & 2.90 $\pm$ 0.08 &    0.13 $\pm$ 0.16 \\
$\mu$~Leo      & K2III    & 4474 $\pm$  60 & 2.51 $\pm$ 0.11 &    0.25 $\pm$ 0.15 \\
$\beta$~Ara    & K3Ib-II  & 4197 $\pm$  50 & 1.05 $\pm$ 0.15 & $-$0.05 $\pm$ 0.39 \\
\hline\noalign{\smallskip}
\multicolumn{5}{c}{Metal-poor dwarfs (MPG)}\\
\hline\noalign{\smallskip}
$\epsilon$~For & K2V*     & 5123 $\pm$  78 & 3.52 $\pm$ 0.08 & $-$0.60  $\pm$ 0.10 \\
HD~107328      & K0IIIb   & 4496 $\pm$  59 & 2.09 $\pm$ 0.13 & $-$0.33  $\pm$ 0.16 \\
Arcturus       & K1.5III  & 4286 $\pm$  35 & 1.60 $\pm$ 0.20 & $-$0.52  $\pm$ 0.08 \\
HD~220009      & K2III    & 4217 $\pm$  60 & 1.43 $\pm$ 0.12 &  $-$0.74 $\pm$ 0.13 \\
\hline
\end{longtable}
}

%% file: Tables/stepar_results_last.tex
\longtab[2]{
\begin{longtable}{cccccccccccc}

\caption{\label{tab:stepar_results} {\scshape StePar} results.}\\
\hline\hline\noalign{\smallskip}
Star & Spectral type & Source & SNR & $T_{\rm eff}$ & $\Delta T_{\rm eff}$ & $\log{g}$ & $\Delta\log{g}$ & $\xi_{\rm micro}$ & $\Delta\xi_{\rm micro}$ & [Fe/H] & $\Delta$[Fe/H] \\
     &               &        &     & [K]           & [K]                  & [dex]     & [dex]           & [km~s$^{-1}$]   & [km~s$^{-1}$]         & [dex]  & [dex] \\
\noalign{\smallskip}\hline\noalign{\smallskip}
\endfirsthead

\multicolumn{12}{c}{Metal-rich dwarfs (MRD)}\\
\hline\noalign{\smallskip}
Procyon        & F5IV-V   & NARVAL        & 765  & 6710 & 50  & 3.96 & 0.09 & 1.70 & 0.06 &    0.07 & 0.03 \\
Procyon        & F5IV-V   & UVES.POP      & 1016 & 6675 & 52  & 3.90 & 0.09 & 1.63 & 0.06 &    0.04 & 0.03 \\
$\beta$~Vir    & F9V      & ESPaDOnS      & 635  & 6206 & 51  & 4.15 & 0.11 & 1.43 & 0.07 &    0.16 & 0.04 \\
$\beta$~Vir    & F9V      & NARVAL        & 400  & 6188 & 44  & 4.12 & 0.09 & 1.33 & 0.05 &    0.19 & 0.03 \\
$\mu$~Ara      & G3IV-V   & HARPS.Archive & 252  & 5814 & 61  & 4.27 & 0.14 & 0.95 & 0.09 &    0.38 & 0.05 \\
$\mu$~Ara      & G3IV-V   & UVES          & 309  & 5889 & 64  & 4.41 & 0.13 & 1.03 & 0.10 &    0.38 & 0.05 \\
18~Sco         & G2Va     & ESPaDOnS      & 383  & 5807 & 53  & 4.40 & 0.13 & 0.69 & 0.11 &    0.12 & 0.04 \\
18~Sco         & G2Va     & NARVAL        & 380  & 5808 & 57  & 4.43 & 0.12 & 0.74 & 0.11 &    0.11 & 0.05 \\
$\alpha$~Cen~A & G2V      & HARPS.Archive & 496  & 5820 & 60  & 4.35 & 0.13 & 0.96 & 0.09 &    0.28 & 0.05 \\
$\alpha$~Cen~A & G2V      & UVES          & 316  & 5824 & 66  & 4.35 & 0.14 & 0.86 & 0.11 &    0.30 & 0.05 \\
Sun            & G2V      & HARPS.Archive & 549  & 5748 & 54  & 4.38 & 0.11 & 0.62 & 0.12 &    0.07 & 0.04 \\
Sun            & G2V      & NARVAL        & 828  & 5766 & 63  & 4.38 & 0.15 & 0.70 & 0.12 &    0.08 & 0.05 \\
$\alpha$~Cen~B & K1V      & HARPS         & 469  & 5088 & 142 & 4.19 & 0.31 & 0.50 & 0.30 &    0.29 & 0.10 \\
$\epsilon$~Eri & K2Vk:    & UVES          & 220  & 5088 & 101 & 4.41 & 0.29 & 0.78 & 0.21 & $-$0.05 & 0.06 \\
$\epsilon$~Eri & K2Vk:    & UVES.POP      & 1653 & 4998 & 100 & 4.40 & 0.23 & 0.41 & 0.28 &    0.00 & 0.06 \\
\hline\noalign{\smallskip}
\multicolumn{12}{c}{Metal-poor dwarfs (MPD)}\\
\hline\noalign{\smallskip}
HD~49933       & F2V      & HARPS.Archive & 319  & 6659 & 41  & 4.05 & 0.10 & 1.54 & 0.05 & $-$0.45 & 0.03 \\
HD~49933       & F2V      & ESPaDOnS      & 1169 & 6640 & 41  & 4.03 & 0.09 & 1.56 & 0.05 & $-$0.43 & 0.03 \\
HD~22879       & F9V      & HARPS.GBOG    & 322  & 5718 & 24  & 3.97 & 0.08 & 1.04 & 0.03 & $-$0.96 & 0.02 \\
HD~22879       & F9V      & NARVAL        & 297  & 5720 & 22  & 3.99 & 0.07 & 1.02 & 0.03 & $-$0.95 & 0.02 \\
$\tau$~Cet     & G8.5V    & ESPaDOnS      & 1238 & 5285 & 90  & 4.46 & 0.22 & 0.43 & 0.21 & $-$0.52 & 0.07 \\
$\tau$~Cet     & G8.5V    & NARVAL        & 357  & 5243 & 138 & 4.37 & 0.34 & 0.50 & 0.35 & $-$0.54 & 0.12 \\
$\mu$~Cas      & G5Vb     & NARVAL        & 269  & 5257 & 66  & 4.32 & 0.15 & 0.42 & 0.24 & $-$0.83 & 0.06 \\
\hline\noalign{\smallskip}
\multicolumn{12}{c}{Metal-rich giants (MRG)}\\
\hline\noalign{\smallskip}
$\beta$~Hyi    & G0V      & HARPS.Archive & 428  & 5825 & 62  & 3.90 & 0.12 & 0.84 & 0.08 & $-$0.02 & 0.05 \\
$\beta$~Hyi    & G0V      & UVES.POP      & 676  & 5824 & 65  & 3.89 & 0.12 & 0.92 & 0.07 & $-$0.05 & 0.05 \\
$\xi$~Hya      & G7III    & HARPS.GBOG    & 391  & 5048 & 71  & 2.87 & 0.23 & 1.17 & 0.07 &    0.20 & 0.06 \\
$\xi$~Hya      & G7III    & ESPaDOnS      & 526  & 5034 & 72  & 2.82 & 0.25 & 1.20 & 0.07 &    0.17 & 0.05 \\
$\epsilon$~Vir & G8III    & ESPaDOnS      & 435  & 5115 & 69  & 2.83 & 0.23 & 1.33 & 0.07 &    0.24 & 0.06 \\
$\epsilon$~Vir & G8III    & HARPS.GBOG    & 392  & 5108 & 74  & 2.85 & 0.23 & 1.33 & 0.07 &    0.24 & 0.06 \\
$\delta$~Eri   & K1III-IV & HARPS.Archive & 525  & 5011 & 95  & 3.60 & 0.23 & 0.77 & 0.13 &    0.19 & 0.06 \\
$\delta$~Eri   & K1III-IV & UVES.POP      & 548  & 4976 & 96  & 3.57 & 0.23 & 0.72 & 0.14 &    0.19 & 0.06 \\
$\beta$~Gem    & K0IIIb   & HARPS.GBOG    & 370  & 4893 & 98  & 2.99 & 0.27 & 1.07 & 0.09 &    0.25 & 0.06 \\
$\beta$~Gem    & K0IIIb   & UVES          & 163  & 4864 & 100 & 2.88 & 0.32 & 1.14 & 0.10 &    0.18 & 0.07 \\
$\mu$~Leo      & K2III    & ESPaDOnS      & 779  & 4518 & 168 & 2.19 & 0.57 & 1.25 & 0.12 &    0.37 & 0.13 \\
$\mu$~Leo      & K2III    & NARVAL        & 402  & 4516 & 166 & 2.24 & 0.55 & 1.34 & 0.13 &    0.32 & 0.12 \\
$\beta$~Ara    & K3Ib-II  & HARPS.GBOG    & 414  & 4413 & 164 & 1.38 & 0.65 & 2.20 & 0.18 &    0.03 & 0.14 \\
\hline\noalign{\smallskip}
\multicolumn{12}{c}{Metal-poor giants (MPD)}\\
\hline\noalign{\smallskip}
$\epsilon$~For & K2V*     & HARPS.GBOG    & 334  & 5092 & 54  & 3.63 & 0.15 & 0.70 & 0.10 & $-$0.53 & 0.04 \\
HD~107328      & K0IIIb   & HARPS.GBOG    & 459  & 4418 & 62  & 1.85 & 0.26 & 1.71 & 0.07 & $-$0.43 & 0.06 \\
HD~107328      & K0IIIb   & NARVAL        & 375  & 4442 & 60  & 1.98 & 0.25 & 1.68 & 0.06 & $-$0.39 & 0.06 \\
Arcturus       & K1.5III  & HARPS.Archive & 475  & 4337 & 60  & 1.87 & 0.25 & 1.64 & 0.07 & $-$0.44 & 0.06 \\
Arcturus       & K1.5III  & UVES.POP      & 1208 & 4338 & 56  & 1.83 & 0.24 & 1.59 & 0.06 & $-$0.44 & 0.05 \\
HD~220009      & K2III    & HARPS.GBOG    & 347  & 4346 & 46  & 1.79 & 0.20 & 1.42 & 0.05 & $-$0.65 & 0.04 \\
HD~220009      & K2III    & NARVAL        & 376  & 4346 & 54  & 1.93 & 0.24 & 1.45 & 0.06 & $-$0.63 & 0.05 \\
\hline
\end{longtable}
}

%% file: Tables/lines_fe_i.tex
\longtab[3]{
\begin{longtable}{ccccccc}

\caption{\label{tab:line_table_all_fe_i} Merged Fe~{\sc i} linelists.}\\
\hline\hline\noalign{\smallskip}
${\lambda}_{\rm air}$ & ${\chi}_{\rm l}$ & $\log{gf}$ & \multicolumn{4}{c}{List}\\
{[}\AA{]}             & [eV]             &            & MRD & MPD & MRG & MPG \\
\noalign{\smallskip}\hline\noalign{\smallskip}
\endfirsthead

\caption{Merged Fe~{\sc i} linelists (cont.).}\\
\hline\hline\noalign{\smallskip}
${\lambda}_{\rm air}$ & ${\chi}_{\rm l}$ & $\log{gf}$ & \multicolumn{4}{c}{List}\\
{[}\AA{]}             & [eV]             &            & MRD & MPD & MRG & MPG \\
\noalign{\smallskip}\hline\noalign{\smallskip}
\endhead

\hline
\endfoot

\hline
\noalign{\smallskip}
\endlastfoot

\noalign{\smallskip}
4808.148 & 3.25 & $-$2.690 & \textbullet &             & \textbullet &             \\
4809.938 & 3.57 & $-$2.620 & \textbullet &             &             & \textbullet \\
4869.463 & 3.55 & $-$2.420 & \textbullet &             & \textbullet & \textbullet \\
4875.877 & 3.33 & $-$1.900 & \textbullet & \textbullet &             &             \\
4877.604 & 3.00 & $-$3.050 &             &             & \textbullet & \textbullet \\
4882.143 & 3.42 & $-$1.480 &             & \textbullet & \textbullet &             \\
4892.859 & 4.22 & $-$1.290 & \textbullet & \textbullet & \textbullet & \textbullet \\
4903.310 & 2.88 & $-$0.903 &             &             & \textbullet &             \\
4905.133 & 3.93 & $-$1.730 & \textbullet &             &             &             \\
4907.732 & 3.43 & $-$1.700 & \textbullet & \textbullet &             &             \\
4917.230 & 4.19 & $-$1.080 &             & \textbullet &             &             \\
4924.770 & 2.28 & $-$2.216 & \textbullet & \textbullet &             &             \\
4939.687 & 0.86 & $-$3.336 & \textbullet & \textbullet &             & \textbullet \\
4946.387 & 3.37 & $-$1.110 & \textbullet & \textbullet & \textbullet & \textbullet \\
4950.105 & 3.42 & $-$1.490 & \textbullet & \textbullet &             &             \\
4961.913 & 3.63 & $-$2.190 & \textbullet &             &             &             \\
4962.572 & 4.18 & $-$1.182 & \textbullet & \textbullet & \textbullet &             \\
4966.088 & 3.33 & $-$0.792 & \textbullet & \textbullet &             &             \\
4969.917 & 4.22 & $-$0.710 &             & \textbullet &             &             \\
4985.253 & 3.93 & $-$0.447 &             & \textbullet &             &             \\
4986.223 & 4.22 & $-$1.290 &             & \textbullet &             &             \\
4992.785 & 4.26 & $-$2.350 &             &             & \textbullet &             \\
4993.680 & 4.21 & $-$1.370 & \textbullet &             &             &             \\
4994.130 & 0.92 & $-$3.058 & \textbullet & \textbullet & \textbullet & \textbullet \\
5002.792 & 3.40 & $-$1.460 & \textbullet & \textbullet & \textbullet & \textbullet \\
5012.695 & 4.28 & $-$1.690 & \textbullet &             & \textbullet &             \\
5014.942 & 3.94 & $-$0.183 &             & \textbullet & \textbullet & \textbullet \\
5022.235 & 3.98 & $-$0.370 &             & \textbullet & \textbullet &             \\
5023.186 & 4.28 & $-$1.500 &             &             & \textbullet &             \\
5029.618 & 3.42 & $-$1.950 & \textbullet &             &             &             \\
5031.914 & 4.37 & $-$1.570 &             &             &             & \textbullet \\
5044.211 & 2.85 & $-$2.038 & \textbullet & \textbullet & \textbullet &             \\
5048.436 & 3.96 & $-$1.005 &             &             & \textbullet &             \\
5049.820 & 2.28 & $-$1.348 &             & \textbullet & \textbullet &             \\
5054.642 & 3.64 & $-$1.921 & \textbullet &             &             & \textbullet \\
5060.078 & 0.00 & $-$5.431 &             & \textbullet & \textbullet &             \\
5067.150 & 4.22 & $-$0.970 & \textbullet & \textbullet &             &             \\
5068.766 & 2.94 & $-$1.041 &             &             & \textbullet &             \\
5074.748 & 4.22 & $-$0.230 &             & \textbullet & \textbullet &             \\
5079.223 & 2.20 & $-$2.068 &             & \textbullet &             &             \\
5079.740 & 0.99 & $-$3.221 &             & \textbullet &             &             \\
5083.338 & 0.96 & $-$2.939 &             & \textbullet & \textbullet &             \\
5088.153 & 4.15 & $-$1.680 &             &             &             & \textbullet \\
5090.773 & 4.26 & $-$0.440 & \textbullet & \textbullet &             &             \\
5104.438 & 4.28 & $-$1.590 &             &             & \textbullet &             \\
5107.447 & 0.99 & $-$3.089 &             & \textbullet &             &             \\
5109.652 & 4.30 & $-$0.980 &             &             & \textbullet &             \\
5127.359 & 0.92 & $-$3.306 &             & \textbullet &             &             \\
5133.688 & 4.18 &    0.360 &             & \textbullet & \textbullet &             \\
5141.739 & 2.42 & $-$1.978 &             &             & \textbullet & \textbullet \\
5143.723 & 2.20 & $-$3.690 &             &             & \textbullet &             \\
5150.839 & 0.99 & $-$3.008 &             & \textbullet & \textbullet &             \\
5151.911 & 1.01 & $-$3.322 & \textbullet & \textbullet &             &             \\
5159.058 & 4.28 & $-$0.820 &             & \textbullet &             &             \\
5162.273 & 4.18 &    0.020 &             & \textbullet &             &             \\
5197.936 & 4.30 & $-$1.540 & \textbullet &             & \textbullet &             \\
5198.711 & 2.22 & $-$2.135 & \textbullet & \textbullet & \textbullet &             \\
5213.806 & 3.94 & $-$2.760 &             &             & \textbullet &             \\
5215.180 & 3.27 & $-$0.861 &             &             & \textbullet &             \\
5216.274 & 1.61 & $-$2.082 &             &             & \textbullet &             \\
5217.389 & 3.21 & $-$1.074 & \textbullet & \textbullet & \textbullet & \textbullet \\
5225.526 & 0.11 & $-$4.789 &             & \textbullet &             &             \\
5228.376 & 4.22 & $-$1.190 &             & \textbullet &             &             \\
5229.845 & 3.28 & $-$0.967 &             & \textbullet &             &             \\
5242.491 & 3.63 & $-$0.967 & \textbullet & \textbullet & \textbullet & \textbullet \\
5243.776 & 4.26 & $-$1.050 & \textbullet & \textbullet & \textbullet & \textbullet \\
5247.050 & 0.09 & $-$4.949 & \textbullet & \textbullet &             &             \\
5250.209 & 0.12 & $-$4.933 &             & \textbullet &             &             \\
5250.646 & 2.20 & $-$2.180 &             & \textbullet &             &             \\
5253.462 & 3.28 & $-$1.579 &             & \textbullet & \textbullet &             \\
5285.127 & 4.44 & $-$1.660 & \textbullet &             & \textbullet & \textbullet \\
5288.525 & 3.70 & $-$1.493 & \textbullet &             &             & \textbullet \\
5293.959 & 4.14 & $-$1.770 &             &             & \textbullet & \textbullet \\
5294.547 & 3.64 & $-$2.760 & \textbullet &             & \textbullet &             \\
5295.312 & 4.42 & $-$1.590 & \textbullet &             & \textbullet & \textbullet \\
5307.361 & 1.61 & $-$2.912 & \textbullet & \textbullet &             & \textbullet \\
5321.108 & 4.44 & $-$1.089 &             &             &             & \textbullet \\
5322.041 & 2.28 & $-$2.802 & \textbullet &             &             & \textbullet \\
5339.929 & 3.27 & $-$0.635 &             &             &             & \textbullet \\
5364.871 & 4.45 &    0.228 &             &             & \textbullet & \textbullet \\
5373.709 & 4.47 & $-$0.710 & \textbullet & \textbullet &             &             \\
5379.574 & 3.70 & $-$1.514 & \textbullet & \textbullet & \textbullet & \textbullet \\
5386.333 & 4.15 & $-$1.670 & \textbullet &             & \textbullet & \textbullet \\
5389.479 & 4.42 & $-$0.410 &             & \textbullet & \textbullet &             \\
5397.618 & 3.63 & $-$2.528 &             &             & \textbullet &             \\
5398.279 & 4.45 & $-$0.630 & \textbullet & \textbullet & \textbullet &             \\
5400.501 & 4.37 & $-$0.160 & \textbullet & \textbullet &             &             \\
5401.266 & 4.32 & $-$1.820 & \textbullet &             & \textbullet &             \\
5409.133 & 4.37 & $-$1.200 & \textbullet & \textbullet &             &             \\
5417.033 & 4.42 & $-$1.580 & \textbullet &             & \textbullet &             \\
5424.068 & 4.32 &    0.520 &             & \textbullet &             &             \\
5436.295 & 4.39 & $-$1.440 & \textbullet &             & \textbullet & \textbullet \\
5436.588 & 2.28 & $-$2.964 &             &             & \textbullet &             \\
5441.339 & 4.31 & $-$1.630 & \textbullet &             & \textbullet & \textbullet \\
5445.042 & 4.39 & $-$0.020 &             & \textbullet & \textbullet & \textbullet \\
5460.873 & 3.07 & $-$3.426 &             &             & \textbullet & \textbullet \\
5461.550 & 4.45 & $-$1.800 & \textbullet &             & \textbullet &             \\
5463.275 & 4.44 &    0.070 &             &             & \textbullet & \textbullet \\
5464.280 & 4.14 & $-$1.402 &             &             & \textbullet &             \\
5466.396 & 4.37 & $-$0.630 & \textbullet & \textbullet &             & \textbullet \\
5470.093 & 4.45 & $-$1.710 & \textbullet &             & \textbullet &             \\
5472.709 & 4.21 & $-$1.495 & \textbullet &             &             &             \\
5473.900 & 4.15 & $-$0.720 & \textbullet & \textbullet &             &             \\
5483.099 & 4.15 & $-$1.392 & \textbullet &             &             & \textbullet \\
5501.465 & 0.96 & $-$3.046 &             & \textbullet &             &             \\
5506.779 & 0.99 & $-$2.795 &             & \textbullet &             &             \\
5522.446 & 4.21 & $-$1.450 & \textbullet & \textbullet &             &             \\
5536.580 & 2.83 & $-$3.710 &             &             &             & \textbullet \\
5539.280 & 3.64 & $-$2.560 &             &             &             & \textbullet \\
5543.147 & 3.70 & $-$1.470 &             & \textbullet &             &             \\
5543.936 & 4.22 & $-$1.040 & \textbullet & \textbullet & \textbullet & \textbullet \\
5546.506 & 4.37 & $-$1.210 &             & \textbullet &             &             \\
5549.949 & 3.70 & $-$2.810 &             &             & \textbullet &             \\
5554.894 & 4.55 & $-$0.270 &             & \textbullet &             &             \\
5560.212 & 4.44 & $-$1.090 & \textbullet & \textbullet & \textbullet &             \\
5572.842 & 3.40 & $-$0.289 &             &             &             & \textbullet \\
5576.089 & 3.43 & $-$0.900 &             & \textbullet & \textbullet & \textbullet \\
5618.632 & 4.21 & $-$1.255 & \textbullet & \textbullet & \textbullet & \textbullet \\
5619.595 & 4.39 & $-$1.600 & \textbullet &             &             & \textbullet \\
5633.946 & 4.99 & $-$0.230 &             & \textbullet &             &             \\
5635.822 & 4.26 & $-$1.790 & \textbullet &             &             &             \\
5636.696 & 3.64 & $-$2.510 & \textbullet &             & \textbullet &             \\
5638.262 & 4.22 & $-$0.720 & \textbullet & \textbullet &             &             \\
5641.434 & 4.26 & $-$1.080 & \textbullet & \textbullet &             &             \\
5649.987 & 5.10 & $-$0.820 & \textbullet &             &             & \textbullet \\
5651.469 & 4.47 & $-$1.900 & \textbullet &             &             &             \\
5652.318 & 4.26 & $-$1.850 & \textbullet &             &             &             \\
5653.865 & 4.39 & $-$1.540 & \textbullet &             &             & \textbullet \\
5655.176 & 5.06 & $-$0.600 & \textbullet &             &             & \textbullet \\
5661.345 & 4.28 & $-$1.756 & \textbullet &             &             & \textbullet \\
5662.516 & 4.18 & $-$0.447 & \textbullet & \textbullet &             & \textbullet \\
5679.023 & 4.65 & $-$0.820 & \textbullet & \textbullet & \textbullet & \textbullet \\
5691.497 & 4.30 & $-$1.450 &             & \textbullet &             &             \\
5696.089 & 4.55 & $-$1.720 & \textbullet &             &             &             \\
5701.544 & 2.56 & $-$2.193 & \textbullet & \textbullet &             &             \\
5705.464 & 4.30 & $-$1.355 & \textbullet &             &             &             \\
5717.833 & 4.28 & $-$0.990 & \textbullet & \textbullet & \textbullet & \textbullet \\
5720.886 & 4.55 & $-$1.631 & \textbullet &             & \textbullet & \textbullet \\
5731.762 & 4.26 & $-$1.200 & \textbullet & \textbullet &             & \textbullet \\
5732.296 & 4.99 & $-$1.460 & \textbullet &             &             &             \\
5741.848 & 4.26 & $-$1.672 & \textbullet &             &             &             \\
5759.262 & 4.65 & $-$2.216 &             &             & \textbullet &             \\
5778.453 & 2.59 & $-$3.430 &             &             & \textbullet &             \\
5784.658 & 3.40 & $-$2.547 &             &             & \textbullet &             \\
5844.918 & 4.15 & $-$3.054 &             &             & \textbullet &             \\
5849.683 & 3.70 & $-$2.890 &             &             &             & \textbullet \\
5852.219 & 4.55 & $-$1.230 & \textbullet &             & \textbullet &             \\
5853.148 & 1.49 & $-$5.180 &             &             & \textbullet & \textbullet \\
5855.076 & 4.61 & $-$1.478 & \textbullet &             & \textbullet &             \\
5856.088 & 4.29 & $-$1.327 &             &             & \textbullet &             \\
5858.778 & 4.22 & $-$2.160 &             &             & \textbullet &             \\
5861.109 & 4.28 & $-$2.304 &             &             & \textbullet & \textbullet \\
5883.816 & 3.96 & $-$1.260 & \textbullet & \textbullet &             & \textbullet \\
5902.473 & 4.59 & $-$1.710 &             &             &             & \textbullet \\
5905.671 & 4.65 & $-$0.690 & \textbullet & \textbullet & \textbullet & \textbullet \\
5909.972 & 3.21 & $-$2.587 & \textbullet &             & \textbullet & \textbullet \\
5916.247 & 2.45 & $-$2.994 & \textbullet &             & \textbullet & \textbullet \\
5927.789 & 4.65 & $-$0.990 & \textbullet &             & \textbullet &             \\
5929.676 & 4.55 & $-$1.310 & \textbullet &             & \textbullet &             \\
5930.180 & 4.65 & $-$0.230 & \textbullet & \textbullet & \textbullet &             \\
5934.654 & 3.93 & $-$1.070 & \textbullet & \textbullet &             & \textbullet \\
5940.991 & 4.18 & $-$2.050 &             &             &             & \textbullet \\
5952.718 & 3.98 & $-$1.340 &             &             &             & \textbullet \\
5956.694 & 0.86 & $-$4.599 & \textbullet & \textbullet & \textbullet & \textbullet \\
6003.011 & 3.88 & $-$1.100 & \textbullet & \textbullet & \textbullet & \textbullet \\
6012.210 & 2.22 & $-$4.038 &             &             &             & \textbullet \\
6019.365 & 3.57 & $-$3.310 &             &             &             & \textbullet \\
6024.057 & 4.55 & $-$0.120 & \textbullet & \textbullet &             & \textbullet \\
6027.051 & 4.08 & $-$1.089 & \textbullet & \textbullet & \textbullet & \textbullet \\
6056.005 & 4.73 & $-$0.320 &             & \textbullet &             &             \\
6065.482 & 2.61 & $-$1.529 &             & \textbullet & \textbullet &             \\
6079.008 & 4.65 & $-$1.020 & \textbullet & \textbullet & \textbullet & \textbullet \\
6082.710 & 2.22 & $-$3.576 & \textbullet &             & \textbullet & \textbullet \\
6093.643 & 4.61 & $-$1.400 & \textbullet &             & \textbullet & \textbullet \\
6094.373 & 4.65 & $-$1.840 &             &             &             & \textbullet \\
6096.664 & 3.98 & $-$1.830 & \textbullet &             &             & \textbullet \\
6098.244 & 4.56 & $-$1.859 & \textbullet &             & \textbullet & \textbullet \\
6120.246 & 0.92 & $-$5.970 &             &             & \textbullet & \textbullet \\
6127.906 & 4.14 & $-$1.399 & \textbullet & \textbullet &             & \textbullet \\
6136.615 & 2.45 & $-$1.402 &             & \textbullet &             &             \\
6136.994 & 2.20 & $-$2.950 &             & \textbullet &             &             \\
6137.691 & 2.59 & $-$1.402 &             & \textbullet &             &             \\
6151.617 & 2.18 & $-$3.295 & \textbullet & \textbullet &             & \textbullet \\
6165.360 & 4.14 & $-$1.473 & \textbullet & \textbullet &             & \textbullet \\
6170.506 & 4.80 & $-$0.440 & \textbullet & \textbullet &             &             \\
6173.334 & 2.22 & $-$2.880 & \textbullet & \textbullet & \textbullet & \textbullet \\
6180.203 & 2.73 & $-$2.591 & \textbullet &             &             &             \\
6187.989 & 3.94 & $-$1.620 & \textbullet &             & \textbullet & \textbullet \\
6191.557 & 2.43 & $-$1.416 &             & \textbullet &             &             \\
6199.506 & 2.56 & $-$4.430 &             &             &             & \textbullet \\
6200.312 & 2.61 & $-$2.433 & \textbullet & \textbullet &             & \textbullet \\
6213.429 & 2.22 & $-$2.481 & \textbullet & \textbullet &             & \textbullet \\
6219.280 & 2.20 & $-$2.432 & \textbullet & \textbullet &             & \textbullet \\
6220.780 & 3.88 & $-$2.058 &             &             &             & \textbullet \\
6226.734 & 3.88 & $-$2.120 & \textbullet &             &             &             \\
6229.226 & 2.85 & $-$2.805 & \textbullet &             & \textbullet & \textbullet \\
6230.722 & 2.56 & $-$1.281 & \textbullet & \textbullet &             &             \\
6240.646 & 2.22 & $-$3.230 & \textbullet & \textbullet & \textbullet & \textbullet \\
6246.318 & 3.60 & $-$0.771 & \textbullet & \textbullet & \textbullet &             \\
6252.555 & 2.40 & $-$1.699 & \textbullet & \textbullet & \textbullet & \textbullet \\
6265.132 & 2.18 & $-$2.550 & \textbullet & \textbullet & \textbullet & \textbullet \\
6270.223 & 2.86 & $-$2.470 & \textbullet &             & \textbullet & \textbullet \\
6271.278 & 3.33 & $-$2.703 &             &             & \textbullet & \textbullet \\
6280.617 & 0.86 & $-$4.390 &             & \textbullet &             &             \\
6290.543 & 2.59 & $-$4.330 &             &             &             & \textbullet \\
6297.793 & 2.22 & $-$2.737 &             & \textbullet &             & \textbullet \\
6301.500 & 3.65 & $-$0.720 &             & \textbullet & \textbullet & \textbullet \\
6311.499 & 2.83 & $-$3.141 & \textbullet &             &             & \textbullet \\
6315.811 & 4.08 & $-$1.630 & \textbullet &             &             & \textbullet \\
6322.685 & 2.59 & $-$2.430 & \textbullet & \textbullet & \textbullet & \textbullet \\
6335.330 & 2.20 & $-$2.177 & \textbullet & \textbullet & \textbullet & \textbullet \\
6336.823 & 3.69 & $-$0.852 & \textbullet & \textbullet & \textbullet & \textbullet \\
6338.876 & 4.80 & $-$0.960 & \textbullet &             &             &             \\
6344.148 & 2.43 & $-$2.919 &             & \textbullet &             &             \\
6355.028 & 2.85 & $-$2.340 &             & \textbullet &             &             \\
6380.743 & 4.19 & $-$1.375 & \textbullet &             &             & \textbullet \\
6393.600 & 2.43 & $-$1.452 &             & \textbullet & \textbullet &             \\
6400.317 & 0.92 & $-$4.318 & \textbullet & \textbullet &             &             \\
6411.648 & 3.65 & $-$0.596 &             & \textbullet &             &             \\
6421.350 & 2.28 & $-$2.012 & \textbullet & \textbullet &             &             \\
6430.845 & 2.18 & $-$2.005 & \textbullet & \textbullet &             &             \\
6469.192 & 4.84 & $-$0.730 & \textbullet & \textbullet &             &             \\
6475.624 & 2.56 & $-$2.941 & \textbullet & \textbullet &             &             \\
6481.870 & 2.28 & $-$2.981 & \textbullet & \textbullet &             & \textbullet \\
6494.980 & 2.40 & $-$1.268 &             & \textbullet &             &             \\
6495.741 & 4.84 & $-$0.840 & \textbullet &             &             &             \\
6496.466 & 4.80 & $-$0.530 & \textbullet & \textbullet &             & \textbullet \\
6498.938 & 0.96 & $-$4.687 & \textbullet &             &             & \textbullet \\
6518.366 & 2.83 & $-$2.438 &             &             &             & \textbullet \\
6533.928 & 4.56 & $-$1.360 & \textbullet & \textbullet & \textbullet & \textbullet \\
6546.238 & 2.76 & $-$1.536 &             & \textbullet & \textbullet & \textbullet \\
6574.227 & 0.99 & $-$5.004 & \textbullet &             &             &             \\
6581.209 & 1.49 & $-$4.679 &             &             &             & \textbullet \\
6591.313 & 4.59 & $-$2.081 &             &             & \textbullet & \textbullet \\
6592.912 & 2.73 & $-$1.473 & \textbullet & \textbullet &             & \textbullet \\
6593.869 & 2.43 & $-$2.420 & \textbullet & \textbullet &             & \textbullet \\
6597.559 & 4.80 & $-$0.970 & \textbullet &             &             & \textbullet \\
6608.025 & 2.28 & $-$3.930 & \textbullet &             &             & \textbullet \\
6609.110 & 2.56 & $-$2.691 & \textbullet & \textbullet &             & \textbullet \\
6627.544 & 4.55 & $-$1.590 & \textbullet &             &             &             \\
6633.412 & 4.84 & $-$1.390 & \textbullet &             &             &             \\
6633.749 & 4.56 & $-$0.799 & \textbullet &             &             &             \\
6648.080 & 1.01 & $-$5.918 &             &             &             & \textbullet \\
6703.566 & 2.76 & $-$3.060 & \textbullet &             & \textbullet &             \\
6710.318 & 1.49 & $-$4.764 & \textbullet &             &             &             \\
6713.743 & 4.80 & $-$1.500 & \textbullet &             & \textbullet &             \\
6716.236 & 4.58 & $-$1.836 & \textbullet &             &             &             \\
6725.356 & 4.10 & $-$2.100 &             &             & \textbullet & \textbullet \\
6750.151 & 2.42 & $-$2.618 & \textbullet & \textbullet & \textbullet & \textbullet \\
6752.707 & 4.64 & $-$1.204 & \textbullet &             &             &             \\
\hline

\end{longtable}
}

%% file: Tables/lines_fe_ii.tex
\longtab[4]{
\begin{longtable}{ccccccc}

\caption{\label{tab:line_table_all_fe_ii} Merged Fe~{\sc ii} linelists.}\\
\hline\hline\noalign{\smallskip}
${\lambda}_{\rm air}$ & ${\chi}_{\rm l}$ & $\log{gf}$ & \multicolumn{4}{c}{List}\\
{[}\AA{]}             & [eV]             &            & MRD & MPD & MRG & MPG \\
\noalign{\smallskip}\hline\noalign{\smallskip}
\endfirsthead

\caption{Merged Fe~{\sc ii} linelists (cont.).}\\
\hline\hline\noalign{\smallskip}
${\lambda}_{\rm air}$ & ${\chi}_{\rm l}$ & $\log{gf}$ & \multicolumn{4}{c}{List}\\
{[}\AA{]}             & [eV]             &            & MRD & MPD & MRG & MPG \\
\noalign{\smallskip}\hline\noalign{\smallskip}
\endhead

\hline
\endfoot

\hline
\noalign{\smallskip}
\endlastfoot

\noalign{\smallskip}
4993.350 & 2.81 & $-$3.684 & \textbullet & \textbullet & \textbullet &             \\
5197.568 & 3.23 & $-$2.220 & \textbullet & \textbullet & \textbullet &             \\
5234.623 & 3.22 & $-$2.180 & \textbullet & \textbullet & \textbullet & \textbullet \\
5256.932 & 2.89 & $-$4.182 & \textbullet &             & \textbullet &             \\
5264.802 & 3.23 & $-$3.130 &             & \textbullet &             &             \\
5284.103 & 2.89 & $-$3.195 &             & \textbullet &             &             \\
5325.552 & 3.22 & $-$3.160 &             & \textbullet &             & \textbullet \\
5414.070 & 3.22 & $-$3.580 &             &             & \textbullet &             \\
5425.248 & 3.20 & $-$3.220 & \textbullet & \textbullet & \textbullet & \textbullet \\
5534.838 & 3.25 & $-$2.865 &             & \textbullet &             &             \\
5991.371 & 3.15 & $-$3.647 & \textbullet &             &             &             \\
6084.102 & 3.20 & $-$3.881 & \textbullet &             & \textbullet &             \\
6149.246 & 3.89 & $-$2.841 & \textbullet &             & \textbullet & \textbullet \\
6238.386 & 3.89 & $-$2.600 &             & \textbullet & \textbullet &             \\
6247.557 & 3.89 & $-$2.435 &             & \textbullet &             &             \\
6369.459 & 2.89 & $-$4.110 & \textbullet &             &             & \textbullet \\
6416.919 & 3.89 & $-$2.877 & \textbullet &             &             &             \\
6432.676 & 2.89 & $-$3.570 & \textbullet & \textbullet & \textbullet &             \\
6456.380 & 3.90 & $-$2.185 &             & \textbullet & \textbullet & \textbullet \\
6516.077 & 2.89 & $-$3.310 & \textbullet & \textbullet &             &             \\
\hline

\end{longtable}
}